# Hybrid-driven Trajectory Prediction Based on Group Emotion

Chaochao Li[1], Mingliang Xu*[1]

## 1 Introduction

Trajectory prediction is a very important research topic in military training, social stability, architectural design, digital entertainment, and other related fields. There are many complex factors that affect movement trajectories of a crowd. The characteristics of crowd movements are different in various scenes. There is an urgent need for a general method to calculate trajectories of crowd, which is more realistic and credible.

Our goal is to propose a general, efficient, and reliable method for crowd trajectory prediction. Because crowd movements are very complex, it is difficult to work on all the scenarios using only one kind of methods. The current trajectory prediction methods are mainly divided into two categories: model driven and data driven. The model driven methods are less realistic, but more controllable, efficient, and universal. The data-driven methods are more realistic, but less controllable, efficient, and universal. It is very difficult to calculate real crowd behaviors, while considering the generality of the method. Most methods often focus on movement modeling of crowd in a special situation or one kind of scenes. In addition, in order to calculate crowd movements in large-scale scenes, many methods divide the crowd into groups[1, 2]. Emotion has an important effect on movements[3]. At present, most crowd behavior calculation models consider individual emotions. Few methods consider emotion of groups and calculate the influence of group emotion on individual movements.

We present a hybrid-driven trajectory prediction method based on group emotion. The data driven and model driven methods are combined to make a compromise between the controllability, generality, and efficiency of the method on the basis of simulating more real crowd movements. A hybrid driven method is proposed to improve the reliability of the calculation results based on real crowd data, and ensure the controllability of the model. It reduces the dependence of our model on real data and realizes the complementary advantages of these two kinds of methods. In addition, we divide crowd into groups based on human relations in society. So our method can calculate the movements in different scales. We predict individual movement trajectories according to the trajectories of group and fully consider the influence of the group movement state on the individual movements. Besides we also propose a group emotion calculation method and our method also considers the effect of group emotion on crowd movements.

The main contributions of this paper are summarized as follows:

(1) We propose a hybrid driven trajectory prediction method, which combines data driven and model driven methods. The possible long-term destinations of the research objects are predicted according to data-driven method, then the trajectories from the current positions to predicted destinations are predicted by model driven method. It no longer relies entirely on real

[1] Corresponding author: Mingliang Xu.

Chaochao Li and Mingliang Xu are with School of Information Engineering, Zhengzhou University, Zhengzhou 450000, China.

data. This hybrid-driven method can predict long-term movement trajectories while many models only focus on the short-term trajectory prediction. Besides, our model can not only calculate more real trajectories, but also improve the adaptability and scalability of the method.

(2) We propose a group division method based on social relations and a calculation method of group emotion. The individuals in a group with similar movements are regarded as a whole and we take them as a research object, which improves computational efficiency. The human relations are fully considered in the group division and the influence of group emotion on individual movement is studied, which is helpful to describe more real crowd movements.

## 2 Related work

This section summarizes the relevant research of the crowd behavior calculation based on group and hybrid modeling.

### 2.1 Crowd behavior calculation model based group

In the process of crowd behavior modeling, group is a very important concept and it is also the research focus. In reference[2], group is a set of several agents with some connection. Generally, there are many groups and independent individuals in a crowd. According to the collectiveness of group members, the group is divided into consistency group and non-consistency group[4]. Consistency group has a common goal and will not be separated. Non-consistency group has no specific goal or only has a temporary goal. The group members will be separated at any time during the movement. Jing Shao et al. [5] divide group into four states according to the collectiveness of the movement of group members: gas, impure liquid, pure liquid, and solid. The collectiveness of solid group is higher than that of liquid group. The collectiveness of gas group is the weakest. According to the leader agents in groups, they can also be divided into groups with leaders and groups without leaders[6].

Some researchers mainly focus on the group movements modeling. Arno Kamphuis and mark H. overmars [7] take groups as changeable rectangles and circles when they calculate trajectories of groups, and use a Group Potential Fields method to control the movements of crowds. This paper points out that the trajectories of a group should keep a certain gap with the surrounding obstacles, that is to say, the distance between any point on the trajectory of a group and the obstacles should be greater than a certain threshold. At the same time, people always want to walk the shortest path. In fact, "big gap" and "short path" are two conflicting indicators [8]. If the distance between the group and the obstacle is very large, the path of the group will be too long. If people choose a shorter path, the distance between the group and the obstacle is very small, or even people collide with the obstacle. In order to balance the path length and the distance between obstacles and groups, Ron Wein and van den Berg [9] propose a method based on a path graph. Users can specify the gap between the trajectory of group and obstacles and generate a path that meets the requirement of the user. Ron Wein, Jur van den Berg, and Dan Halperin [10] integrate path length and gap information into objective function and calculate an approximate optimal path. Bisagno et al. [1] predict the group trajectory in crowded scenes based on the social LSTM network. If two individuals belong to the same group, they are regarded as a whole when calculating the social pooling layer, and they are not calculated separately to improve the calculation efficiency. Musse and thalmann et al [11] consider the influence of human personality on group behaviors and the group behaviors they study include following, attraction, exclusion, separation, and so on. Eunjung Ju et al. [12] simulate the overall movement characteristics of a group based on the formation information of groups in real videos. Yan Mao et al. [13] study emotional contagion among groups and discuss the effect of emotional contagion on the crowd movements in emergencies. Although this method discusses emotions, it does not discuss the emotional state of the whole group. To sum up, current methods rarely discuss emotions of the whole group, which has a great impact on group movements. Although some works discuss emotional contagion among groups, they do not take the group as a whole and they still focus on the emotion of individuals. Our method not only considers the emotional state of groups, but also studies its influence on movements.

The method taking group as research object cannot describe the movements of small-scale crowd scene in detail. Therefore, we also need to pay attention to the movements of individuals in a group. The relative positions of members in a group are not fixed. However, in real life, small groups are a very important part of crowd and they generally include two to five people. Generally speaking, the greater the number of group members, the distance between these members will increase. They will tend to separate. So that large groups will become small groups [14]. Moussaid [14] and Peters [15] et al. define the number of group members as 2 to 4 by observing the real crowds, and study the movements of each group member. They discuss the relationship between formation and crowd density when the number of group members is 2, 3, and 4. The actual formation of a group is obtained by changing the basic formation. Our model takes group as the research object and proposes a group division method based on emotional contagion. Besides the influence of the human relations on the group division is studied and a group emotion calculation method is proposed according to group movements. We not only study the movements of groups, but also the movements of each group member.

## 2.2 Crowd behavior calculation model based on hybrid modelling

At present, crowd behavior models based on hybrid idea mainly combine macro and micro methods. Kiran Ijaz et al. [16] point out that a large range of crowd behaviors and individual behaviors are both needed to consider. Muzhou Xiong et al. [17] combine the macro model with the micro model to partition the crowd according to the movement characteristics. Each partition runs macro model and micro model independently, and transfers data at the boundary of each partition. Tisera et al. [18] use macro and micro technologies to simulate crowd movements based on the idea of stratification, which can perform global path planning, local collision avoidance and other crowd behaviors at different layers. Xiong et al. [19] propose a hybrid method based on sequence to integrate macro and micro models. The macro model learns the crowd movement patterns and the micro model focuses on the analysis of individual behaviors. Haiying Liu et al [20] propose a movement simulation framework of mixed crowd, which can automatically identify individuals and groups. This method can explicitly trigger the avoidance behaviors of individuals to groups. Murat Haciomeroglu et al. [21] cluster agents according to their positions and velocity using the parallel processing ability of GPU, and these clustering information can be used to perceive the macro information. Agents make decisions based on the surrounding scene information and the far-reaching global information, and determine the path to the destination. Based on these clustering information, agents can reach the destination quickly.

Most of the hybrid modeling methods are the combination of macro models and micro models, and rarely the integration of data-driven and model driven methods. Our model makes full use of the advantages of these two kinds of methods. We use data-driven method to predict the long-term destinations of individuals. Then model-driven method is used to predict the complete trajectories according to the predicted destinations. Our method solves the problem that the data-driven method excessively relies on the real data. It makes up for the deficiency that many prediction methods can only be applied to specific scenarios.

## 3 Overview of our approach

We propose a hybrid driven trajectory prediction method based on group emotion. According to the movement collectiveness, we divide the crowd in a scene into groups and the emotion values of the groups are calculated. A group is regarded as a whole. According to the known historical trajectories of the groups and scene information, we predict their future trajectories. For a predicted object, we search an individual with similar position and movement direction from the database. We take the destination of the similar individual as the destination of the predicted object. At the same time, we also consider their own historical trajectory information. We can get several possible destinations. The positions of all the groups and the current scene information are input into a model-driven method. The model-driven method are used to predict the candidate trajectories from the current positions to the predicted destinations. Finally, according to the predicted trajectories of a group, the movement trajectories of all the group members are calculated by the emotion value of the group (Fig. 1).

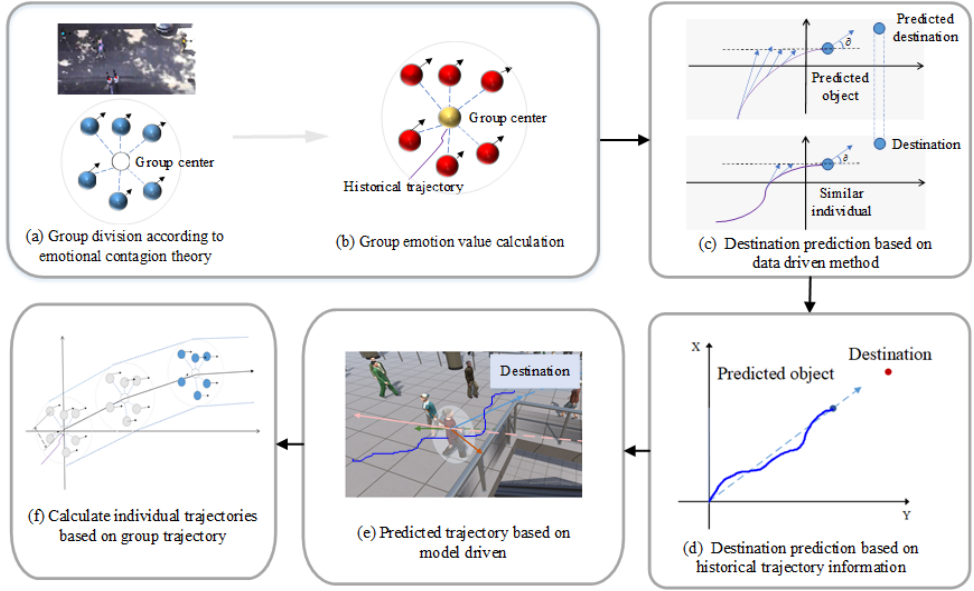

Fig. 1 Overview of our algorithm

### 3.1 Group division according to emotional contagion theory

According to the collectiveness defined in reference[22], the set of agents with high collectiveness (movement similarity) is regarded as a group. These agents often have the same or similar destinations. In this paper, a group is regarded as a whole (as a special agent) and the center of the group is calculated. The center of the group is the representative of the whole group. Its position is obtained by averaging all the positions of the group members. The center positions at different time steps constitute the movement trajectory of group. As shown in Figure 2 is the schematic diagram of a group and the blue circles represent the group members, who are ordinary agents. The black arrows indicate the movement directions of agents. The movement directions of the agents in the same group are similar. The white circles represent the group center. There are many factors that lead to the movement similarity. The special interpersonal relationship of real life (such as relatives, friends, family members, colleagues, etc.) is a very important influencing factors. In addition, through emotional contagion, people have similar emotional state, which leads to similar movements. Because of emotion or interpersonal relationship, the group members cooperate with each other to keep their movements consistent. Individuals who are randomly grouped may have similar destinations in a short period of time. Therefore, we mainly study this kind of stable groups based on emotion and interpersonal relationship.

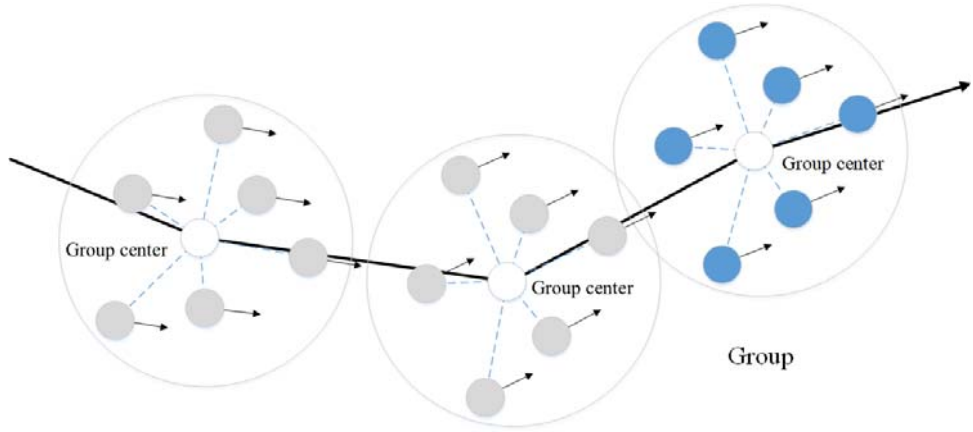

Fig. 2 Schematic diagram of group

Next, we introduce the detailed group division method. People with friends, colleagues, and family members often show the movement similarity, which is a very common phenomenon [23-25]. We divide individuals into groups based on these special relationships. Analyze the intimacy of all the individuals according to their known historical trajectories in the database. Intimacy refers to the degree of movement similarity between individuals due to special interpersonal relationships (for example, friends, couples, family members, etc.). According to the common sense, the intimate distance of people is 0.15m-0.45m and the personal distance is 0.45m-1.2m[26]. People with close relationships will keep within intimate distance, such as family, husband, and wife. Friends and acquaintances will keep within personal distance. Based on this conclusion, we set two thresholds: $T_1$ and $T_2$. If the distance between two individuals from beginning to end is less than a small threshold $T_1$, their intimacy is defined as 1, indicating that they are very close. If the distance between them is greater than the threshold value $T_1$ and less than the threshold value $T_2$, their intimacy is defined as 0.5, indicating that they are generally intimate. Otherwise, we define their intimacy as 0, which means they are not close. Intimacy can be defined as in Equation 1.

$$\text{Inti}(i,j) = \begin{cases} 1 & dis(i,j) \leq T_1 \\ 0.5 & T_1 < dis(i,j) \leq T_2 \\ 0 & dis(i,j) > T_2 \end{cases} \qquad (1)$$

Where $\text{Inti}(i,j)$ represents the intimacy between individual i and individual j, $dis(i,j)$ represents the distance between individual i and individual j. $T_1 = 0.45$, $T_2 = 1.2$. Individuals with intimacy of 1 and 0.5 are considered as a group. The more intimate they are, the stronger their group is, and the more similar their movements are. Figure 3 describes the intimacy. It is represented by a graph. Each node represents an individual in the crowd and the edge $e(i,j) \epsilon E$ represents the intimacy of individual i and individual j.

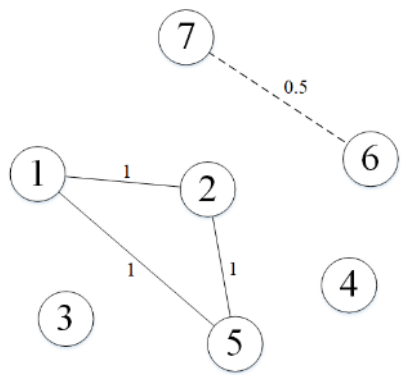

Fig. 3 Intimacy graph

It has been proved that emotional contagion lead to crowd similar movements [27, 28]. Similar emotions are more conducive to group formation [27]. Based on this point of view, we propose a group division method based on emotional contagion, which divides the individuals with similar movement into groups due to the emotional contagion mechanism, that is, the individuals with similar emotion is regarded as a group. In real life, happiness, sadness, and other emotions of friends or family members are more likely to cause people the same emotions. So it is much easier for individuals with special relationships to spread emotions[24].

Negative emotions are more likely to lead to emotional contagion than positive emotions [28]. In general situations, negative emotions, such as panic and fear, are often very weak, and individuals with special relationship can cause emotional contagion. In emergencies (such as explosions, earthquakes, etc.), panic, fear and other negative emotions of people will increase suddenly, and emotional contagion among strangers will become much easier. Individuals with high degree of intimacy are more likely to lead to emotional contagion than those with zero intimacy. So people with higher intimacy often choose to escape together.

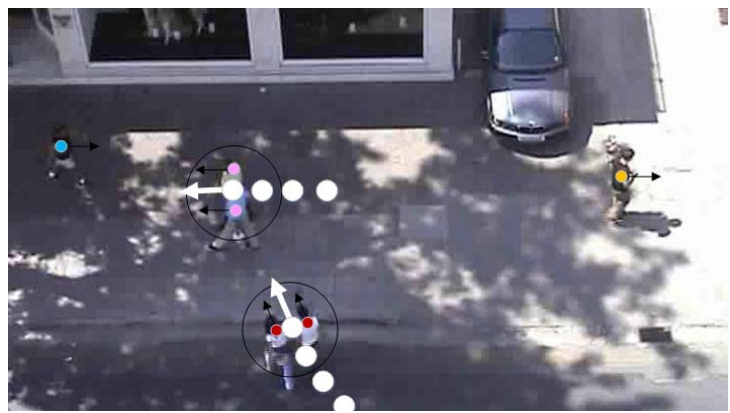

Fig. 4 Schematic diagram of group division based on emotional contagion

As shown in Figure 4, it is a schematic diagram of group division based on emotional contagion. People with special interpersonal relationship are marked with different colors and it is also the final grouping result. The two people marked with green circles may be friends or family members. They always have linguistic communication. Through linguistic communication, emotional contagion occurs between them. Their similar emotions lead to similar movements, which is divided into a group. The two people with red circles are also divided into a group. The average position of all group members is the group center. The trajectory of the group center is the movement trajectory of the whole group. The white points represent the group centers and the trajectories of the white points are the trajectories of the groups.

Next, according to the divided groups, we calculate the emotions of groups. The cohesion of a group is essential indicator of the emotional state [29]. The cohesion of a group refers to the similarity of its members [30]. A small group tends to show strong cohesion. However, with the increase of the number of group members, the differences of behaviors among the group members are highlighted and the cohesiveness is reduced. Therefore, smaller groups tend to have higher cohesion, while larger groups tend to have lower cohesion [31]. In this paper, the collectiveness of individual movements, the size of speed, and the number of group members are used to measure the cohesion of groups, and then the emotional value of groups (that is, the intensity of emotions) is inferred. The detailed calculation equation is as follows.

$$k = 1 + \frac{1}{n(n-1)} \cdot \sum_{i,j \in N \text{ and } i \neq j} \frac{v_i \cdot v_j}{\|v_i\|_2 \|v_j\|_2} - \frac{1}{n(n-1)} \cdot \sum_{i,j \in N \text{ and } i \neq j} \left| \|v_i\|_2 - \|v_j\|_2 \right| - n \quad (2)$$

$$E = \frac{1}{1+e^{-k}} \quad (3)$$

Where n denotes the number of group members, $N$ represents the set of group members, $v_i$ is the velocity of individual i, and E is the emotion value of the group.

## 3.2 Hybrid-driven trajectory prediction

In this section, a hybrid-driven method is used to predict the future movement trajectories. Firstly, the group division method based on emotional contagion in Section 3.1 is used to obtain the groups in the predicted scene. The emotion values of groups are calculated. According to the historical trajectory of a group, the average movement direction is calculated. We search an individual in the database, who have similar position and direction. Next, the destination of the similar individual is taken as the destination of the predicted object. Then, the locations of destinations, the locations of current groups, and scene information are taken as the input of the model-driven method. Finally, these prediction trajectories are optimized according to the scene heat map and probability distribution map of pedestrian speed.

According to the group division method based on emotional contagion in Section 3.1, the groups of the predicted scene is calculated, all the group members are found out, and the centers of the groups are obtained. According to the known trajectories of a group center, the average movement direction of the current position is calculated, as shown in Figure 5. In the database, we search k individuals in similar position and average movement direction, and take k destinations of the similar individuals as the possible destinations of the predicted object. This is a measure of trajectory similarity based on scene semantic information. In real life, people may have similar destinations, who have roughly similar positions and movement directions for a period of time. For example, now a student comes out of the classroom. According to the historical trajectory data of all the students, he or she may go to the canteen, dormitory, and school gate next. According to the historical trajectory of the research object, it continues to move along the current moving direction to get a candidate destination. So we can get k+ 1 candidate destinations.

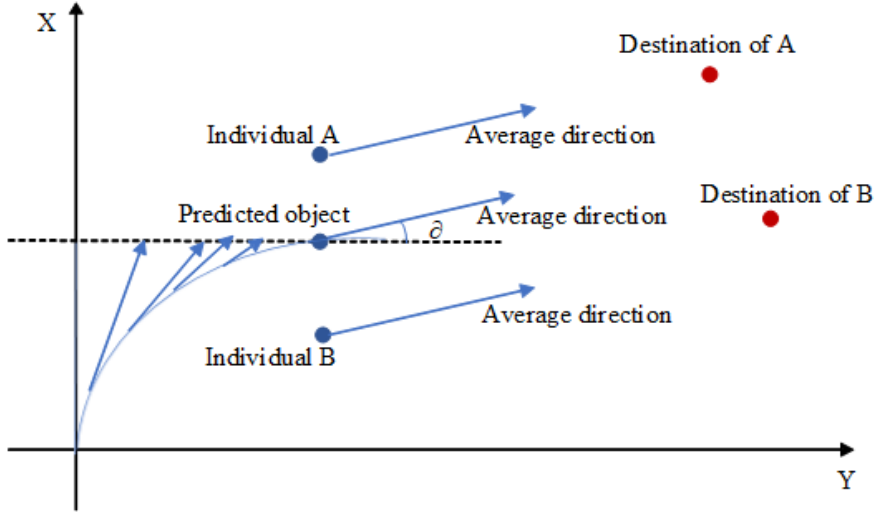

Fig. 5 Schematic diagram of trajectory similarity comparison

Next, we will introduce how to calculate the average movement direction of an individual. The position at time step t of individual i is $(x_t, y_t)$. The position at time step k is $(x_k, y_k)$. The average movement direction of individual i at time step t is defined as follow.

$$(x^t, y^t) = \frac{1}{t-1}\sum_{k=1}^{t-1}(x_t - x_k, y_t - y_k) \qquad (4)$$

According to the predicted destinations, we calculate predicted trajectories using the model-driven method. The model-driven method[32] is used to calculate the trajectories of the predicted objects from the starting points to the destinations. The positions of all the groups in the current scene, their destinations, their speeds, and the position information of all the obstacles are input into the model-driven method to calculate their future movement trajectories.

$$P^{n+1} = f(P^n, Des, V) \qquad (5)$$

Where $P^n$ denotes the locations of all the groups at time step n, $Des$ is the destinations of all the groups, and V is the speed of all the groups. According to the known historical moving data of the research object, the movement speed is calculated.

Based on the above method, we get the movement trajectories of groups. Then we will calculate the movement trajectories of all the group members according to the movement trajectory of group. Group members usually have similar movement trajectories for some time. The degree of similarity and deviation of their trajectories is related to the cohesion of the group (group emotion). When the emotion value of a group is 1 (the group has strong cohesion), it means that the relative position of group members does not change in the movement. When the emotion value of a group is 1, we assume that the positions of the group members at different time steps are standard positions. When the emotion value is less than 1, we calculate movement trajectories of group members according to emotion value of group by the following.

$$P = P_1 + (1 - E) \cdot \partial \qquad (6)$$

Where P denotes the predicted trajectory, $E$ denotes the emotion value of the predicted group, $P_1$ denotes the standard positions, and $\partial$ denotes deviation of position at different time step.

# 4 Experiment

This section evaluates our method. We evaluate our model on three publicly available datasets: ETH[33], UCY[34], and Grand Central Train Station in New York city[35]. We use the following two metrics to report the prediction error. Minimum Average Displacement Error Given K Predictions (minADEK): we choose the closest one from k prediction results to measure its average error. Minimum Final Displacement Error Given K Predictions (minFDEK): we choose the closest predicted destination to measure the error from the k predicted destinations. The destination means the point at the final prediction time instant. Furthermore, we compare the results with the Emotion-Based method (abbreviated as EBM)[32]. Our method is optimized on the basis of the EBM method. The model driven module in our method adopts the EBM method. Based on the EBM method, we use the hybrid driving method and group emotion. To compare our method with other method in a fair manner, we summarize the parameters we used in the following table.

Table 1  List of parameter values we used in our method

| Parameters | Sign | values |
| --- | --- | --- |
| The known time | knowntime | 30 time steps (12s) |
| The prediction time | predicttime | 30 time steps (12s) |
| The number of candidate destinations | k | 5 |
| The radius of a person | r | 0.3m |
| The time of a time step | mytimestep_dur | 0.3999s |
| The time of the last time step of the known trajectory | endtime | --- |
| The neighbourhood range in the model driven method | neighborhoodRange | 10m |
| The mass of a person | m | 60kg |

In the following table, we summary the number of people and the number of groups in the datasets.

Table 2  The number of people and groups in the dataset

| Dataset | Central Train Station | ETH | HOTEL | ZARA01 | ZARA02 | UNI |
| --- | --- | --- | --- | --- | --- | --- |
| The number of people | 12684 | 367 | 420 | 148 | 205 | 118 |
| The number of groups | 51 | 54 | 53 | 8 | 7 | 5 |

We test our method on the datasets. The following table show the values of ADE/FDE.

Table 3  The value of ADE/FDE in Grand Central Train Station in New York city

| Endtime=300 | Endtime=1000 | Endtime=2000 | Endtime=3000 | Endtime=4000 | Endtime=5000 | Endtime=6000 | Endtime=7000 |
| --- | --- | --- | --- | --- | --- | --- | --- |
| 3.341660 /2.737101 | 3.418668 /5.313081 | 3.490345 /5.400639 | 4.359193 /3.874613 | 3.746189 /3.030375 | 3.844779 /4.878712 | 3.898211 /3.334867 | 4.475276 /6.267000 |

Table 4　The value of ADE/FDE in ETH

| Endtime= 8490 | Endtime= 9410 | Endtime= 9930 | Endtime= 10110 | Endtime= 12050 | Endtime= 12070 |
|---|---|---|---|---|---|
| 3.914150/ 6.881780 | 3.802774/ 1.928419 | 5.384488/ 6.621699 | 3.201202/ 0.641327 | 4.266672/ 1.278281 | 3.831911/ 0.941554 |

Table 5　The value of ADE/FDE in HOTEL

| Endtime= 1000 | Endtime= 1100 | Endtime= 1200 | Endtime= 1300 | Endtime= 1500 | Endtime= 1600 |
|---|---|---|---|---|---|
| 1.959727/ 0.044721 | 0.429203/ 0.237697 | 1.746547/ 0.174642 | 1.073894/ 0.202237 | 1.713824/ 0.341823 | 0.087948/ 0.167631 |

Table 6　The value of ADE/FDE in ZARA01

| Endtime= 300 | Endtime= 600 | Endtime= 900 | Endtime= 1200 | Endtime= 1500 | Endtime= 1800 | Endtime= 2100 | Endtime= 7370 |
|---|---|---|---|---|---|---|---|
| 3.283501/ 3.804544 | 0.918685/ 1.423998 | 2.981999/ 0.514924 | 1.488363/ 1.140696 | 2.364243/ 3.415694 | 1.607158/ 3.679737 | 3.302675/ 1.524686 | 2.421865/ 1.232832 |

Table 7　The value of ADE/FDE in ZARA02

| Endtime= 2000 | Endtime= 3000 | Endtime= 4000 | Endtime= 5000 | Endtime= 6000 | Endtime= 7000 | Endtime= 8000 | Endtime= 9000 |
|---|---|---|---|---|---|---|---|
| 2.298538/ 2.785807 | 2.063060/ 1.681055 | 2.619309/ 7.238682 | 2.456768/ 0.406156 | 1.833825/ 0.420587 | 2.505196/ 0.910090 | 2.276266/ 0.354861 | 1.125958/ 0.217961 |

Table 8　The value of ADE/FDE in UNI

| Endtime= 1000 | Endtime= 2000 | Endtime= 3000 | Endtime= 4000 | Endtime= 5000 | Endtime= 6000 | Endtime= 7000 |
|---|---|---|---|---|---|---|
| 2.118697 /-- | 0.753803 /-- | 8.206246 /-- | 4.135471 /-- | 6.279577 /-- | 4.003417 /-- | 2.892705 /-- |

In Fig. 6 and Fig. 7, we qualitatively evaluate the performance of our method in the scene from Grand Central Train Station in New York city. The red trajectories represent the real trajectories and the blue trajectories represent our prediction trajectories. Our prediction result is very close to the real result.

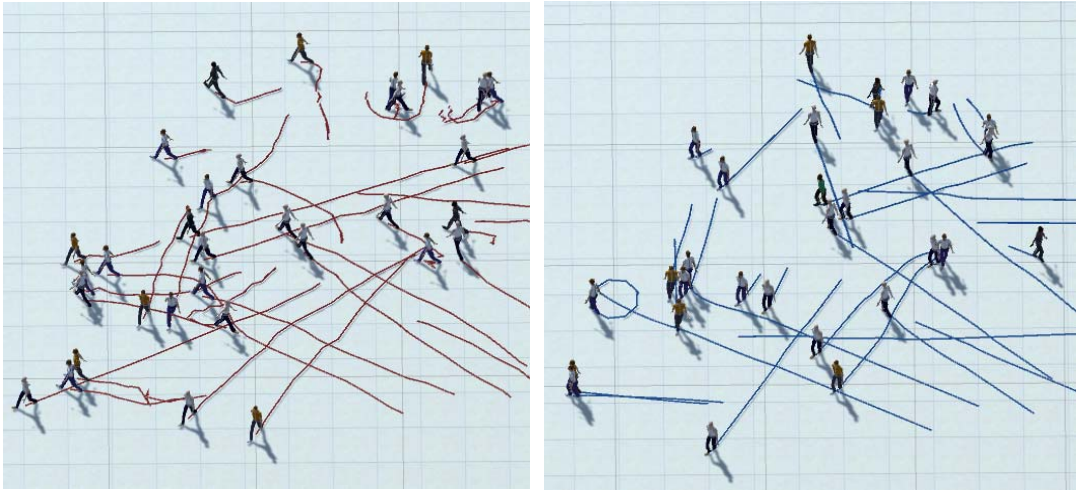

(a) Ground truth          (b) Our prediction result
Fig. 6  Qualitative evaluation of our algorithm in a Scene with fewer people

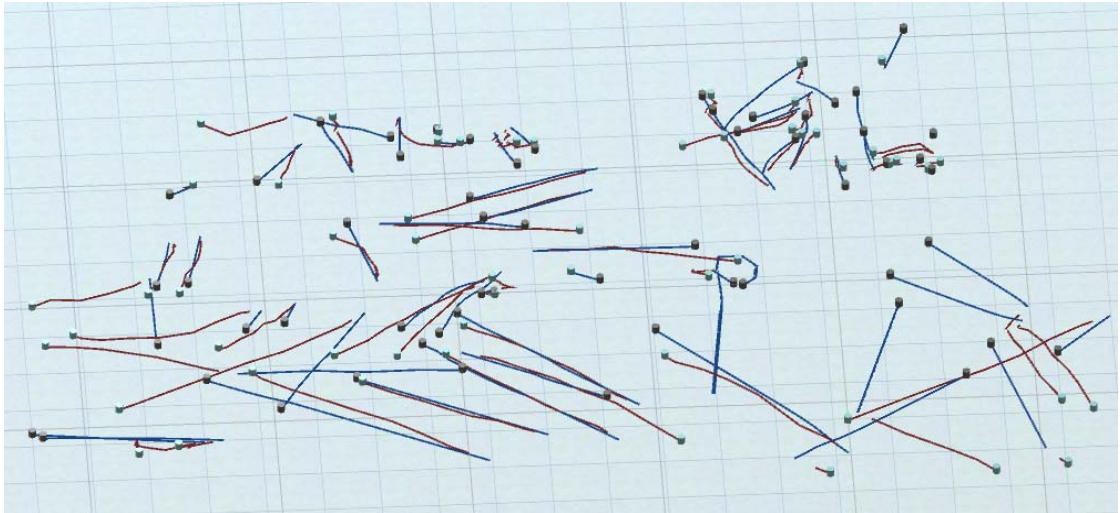

Fig. 7  Qualitative evaluation of our algorithm in a Scene with more people

# 5 Conclusion and limitations

We present a trajectory prediction model based on data and model driven method. The method makes a compromise between controllability, generality, and efficiency, which realizes the complementary advantages of these two kinds of methods. At the same time, the long-term destinations of the research objects are predicted based on the data-driven method. Then, the model driven method is used to calculate the specific trajectories. The long-term movement trajectories of the research objects are predicted. We take groups as research objects and combine macro and micro models. The group is regarded as a whole to predict the movement trajectories macroscopic angle. We predict the movement trajectories of individuals from microcosmic angle based on the group emotion. In addition, this paper proposes a group division method based on emotional contagion, which transfers the research object of emotion from individual to group. At the same time, a group emotion calculation method is proposed and we focus on the effect of group emotion on individual movements. Experiments show that our prediction results are very close to the real results.